
\input harvmac.tex
\def\overarrow#1{{\buildrel #1\over \longrightarrow}}

\def\bar{\overline}

\def\demi{{1\over 2}}

\def\ie{\hbox{\it i.e.}}

\def\eg{\hbox{\it e.g.}}

\def\vev{\hbox{\it vev}}
\Title{\vbox{\baselineskip12pt\hbox{BUHEP-95-tba}\hbox{hep-th/9507082}}}
{\vbox{\centerline{Binary Cosmic Strings}\vskip 2pt}}
\vskip .5in
{\parskip=0pt\baselineskip=12pt
\centerline{ Ryan Rohm {\it and} Indranil Dasgupta}
\bigskip
\centerline{\sl Physics Department}
\centerline{\sl Boston University}
\centerline{\sl Boston, MA  02215}\footnote{}
{\it e-mail: rohm@ryan.bu.edu, dgupta@budoe.bu.edu}}
\vskip .5in

{ The properties of cosmic strings have been investigated in detail
for their implications in early-universe cosmology.  Although many
variations of the basic structure have been discovered, with implications
for both the microscopic and macroscopic properties of cosmic strings,
the cylindrical symmetry of the short-distance structure of the string
is generally unaffected.  In this paper we describe some mechanisms
leading to an asymmetric  structure of the string core, giving the defects
a quasi-two-dimensional character. We also begin to investigate
the consequences of this internal structure for the microscopic
and macroscopic physics.}

\Date{July, 1995}
\newsec{Introduction}

Since the paper of Nielsen and Olesen\ref\NielOle{H.B. Nielsen and
P. Olesen, {\it Nucl.Phys.} {\bf B61} (1973) 45.} describing the vortex
solutions of the Abelian Higgs model, many variations of this
structure have been studied, giving rise to diverse phenomena.
Superconducting cosmic strings \ref\SCS{E. Witten,
{\it Nucl.Phys.} {\bf B249} (1985) 557.}  are a dramatic example of this.
Cosmic strings arise as classical solutions of many Grand Unified Theories
(GUTs) and other models of fundamental interactions.  They may play
an interesting role in cosmology since they are formed in the early
universe, when the order parameter of spontaneous symmetry breaking is
uncorrelated at large distances; later, they may perturb
the space-time metric\ref\metperta
{Ya.B. Zel'dovich, {\it Mon.Not.Roy.Astron.Soc.} {\bf 192}(1980) 663}
\ref\metpertb{A. Vilenkin, {\it Phys.Rev.Lett.} {\bf 46} ( 1981)
1169,1496(E); {\it Phys. Rev.} {\bf D24} (1981) 2082.}\
and contribute to the density fluctuations influencing galaxy formation.
In most previously studied variations
of the basic Nielsen-Olesen vortex solution the cylindrical symmetry is
preserved. We explore another variation in the basic
structure of cosmic strings arising in some non-minimal extensions of
the standard model, in which additional fields appear in the classical
equations and the resulting configurations break the azimuthal symmetry.
(These additional fields may arise, for instance, in an
extended Higgs sector, or may be part of the matter sector of the
theory.) We will discuss two classes of models: one in which pairs
of string are confined due to a spontaneously-broken discrete gauge
symmetry, and another class in which independently stable cosmic strings
form bound pairs of nearly parallel strings because of short-range
attractive forces.

Although there are several mechanisms for obtaining strings with an
asymmetric core, the macroscopic classical motion of the strings can be
described by a geometrical effective action, and only the relative size
of the couplings depends on the details of the short-distance structure.
For this reason, we discuss here the conditions for obtaining
binary strings, and report on the effect of this additional structure
on the macroscopic behavior of the cosmic strings in a separate
paper\ref\strmech{I. Dasgupta and R. Rohm, ``Binary String Dynamics,"
{\it in preparation.} }.

A simple physical argument will serve to illustrate the necessary
condition for a bound state to occur.  Recall
\ref\Bog{E.B. Bogomol'nyi, {\it Sov.J.Nucl.Phys.} {\bf 24} (1976) 449.}
that in the Nielsen and Olesen solution, two vortices
(that is, cross-sections of a cosmic string) will attract
or repel each other depending on the relative strength
of the electric charge and the scalar self-coupling.
For $\beta\equiv({m_S\over m_V})^2<1$, two $n=1$\ vortices have a
short range attraction, so  that the ground state
in the two-vortex sector is the $n=2$ \ vortex.
For $\beta>1$, the magnetic force dominates, leading to a repulsive
short range force.  (The critical \Bog\ case $\beta=1$\ has additional
symmetries, but it will not concern us here.)
In the case of GUT cosmic strings, we will show that
for some ranges of couplings the force between vortices
is attractive at long distances but at short distances
the force becomes repulsive.  This can easily happen, for instance,
when the longer-range force is confining, that is, when
the discrete symmetry (the holonomy of a single string)
is spontaneously broken.  This leads to a long-range force
confining pairs of vortices; if the short-range
interaction is repulsive ($\beta>1$), they do not coalesce into a single
rotationally symmetric vortex.  Such a structure can also arise when the
balance between short-range repulsive forces and short-range attractive
forces between two vortices has a local minimum, leading to a finite
binding energy per unit length.  In either case the classical ground
state consists of a pair of finitely-separated vortices, or in the
(3+1)-dimensional case, a bound pair of cosmic strings.  We will then
discuss examples of both realizations of binary cosmic strings,
beginning with the confining case.

\newsec{Confined binary strings}

We will present two models in which binary strings
arise because of long-range confining forces.
In both cases we have `unary' cosmic strings formed at one scale, which
are charged under a discrete {\it gauge} symmetry.  At a lower scale,
another scalar field breaks this discrete symmetry, and the original
strings become boundaries of domain walls.  These domain walls serve to
bind together pairs of the original strings into binary strings, which
are then in a sense quasi-one-dimensional domain walls, whose width and
thickness are comparable in size.  Later in the paper we will describe
some GUT realizations of such models, such as the
domain-wall-connected strings previously discussed in the literature
\ref\KLS{T.W.B. Kibble, G. Lazarides and Q. Shafi,
{\it Phys.Rev.} {\bf D26} (1983) 435.}.

In the first model, we consider a spontaneously broken $U(1)$ gauge
theory with two charged scalars $\phi$ and $\chi$;
the field $\phi$ has twice the charge of $\chi$.
The Lagrangian is:
\eqn\eone{
L= - {1\over 4}F_{\mu\nu}F^{\mu\nu} + |D_1\phi|^2 +
|D_2\chi|^2 - V.
}
\noindent
where $D_{1\mu} \phi = (\partial_\mu-ieA_\mu)\phi $,\
$D_{2\mu} \chi = (\partial_\mu-\half ieA_\mu)\chi$\ and
$$
V= {\lambda_1\over 8}(|\phi|^2 - v_1^2)^2
+ {\lambda_2\over 8}(|\chi|^2 -  v_2^2)^2
+ \kappa(|\phi|^2-v_1^2)(|\chi|^2 - v_2^2).
$$
We also define
\eqn\eonea{
\gamma = {v_2^2 \over v_1^2}
}

We shall show that for a range of values of $ \gamma $ and the coupling
constants $\lambda_1$ and $\kappa$, and for all values of $\lambda_2$,
the lowest-mass stable configurations are binary strings.  The
finiteness of the mass per unit length of local strings (in contrast to
that of global strings) derives from the proper long-range behavior of
the gauge field; to ensure this, the winding number $\nu_\phi$ of the
phase of $\phi$ must equal twice the winding number of the phase of :
$\nu_\phi = 2 \nu_\chi$.  The smallest finite-mass solution then has
$\nu_\phi=2$.  In the absence of $\chi$ there would be stable
$\nu_\phi=1$\ strings, but in this model separating the $\nu_\phi=2$\
string into two
of the original strings with $\nu_\phi= 1$ creates a domain wall, and for
large separations requires a linearly-increasing energy.
We then need to determine the form of the $\nu_\phi = 2$\ string.
We will demonstrate that for the range of parameters mentioned above,
the cylindrically symmetric configuration is an unstable solution of the
classical field equations; since we have
argued that the string cannot separate, this indicates that the  binary
string is the lowest-energy classical solution.

We will first demonstrate the instability of $\nu_\phi >1$ strings (and
hence existence of binary strings) when we set the coupling $\kappa=0$,
and later demonstrate that this restriction is inessential.
To start, consider the $(\phi , A)$ system if
we totally ignore (\ie, decouple) the field $\chi$; then
$\beta=({m_S\over m_V})^2 = ({\lambda_1\over 4e^2})$, and for
convenience we define $ \alpha = \beta - 1 $.
In that case the proof proceeds along the lines of \Bog, where it was
shown that cylindrically symmetric $n=2$ strings in the theory with
$\alpha>0$ (and only one Higgs field, \ie\ here we ignore $\chi$), are
unstable against specific perturbations which correspond to the
splitting of the string.
The instability occurs because for $\alpha>0$\ there are
asymmetric perturbations around the cylindrically symmetric solution
that lower the energy;
for $\alpha=0$ these asymmetric perturbations of the cylindrically
symmetric classical solutions
are zero modes of the Hamiltonian.

If we make the same perturbations and include the effects of the
lighter Higgs field  $\chi$, (so $\lambda_2$\ and $\gamma$\ are
nonzero), there will be additional positive contributions to the energy,
and the line of instability does not start at $\alpha=0$\ ($\lambda_1 =
4 e^2$).  This is because while energy from the $(\phi , A)$ sector does
not change, the perturbations in A will affect derivative couplings
between A and $\chi$, giving a positive contribution to the energy.
Nevertheless, for any $\alpha>0$\ the strength of the instability is finite
for $\lambda_2=\gamma =0$, and so by continuity there will be some
range of $\lambda_2$ and $\gamma$\ near zero for which the symmetric
vortex is unstable.

We begin by describing the perturbations.
The cylindrically symmetric $\nu_\phi=n$\ solution will have two
independent functions which we call $\Phi$ and $v$\Bog:

\eqn\ethr{
\phi(r,\theta) = \Phi(r)exp(in\theta)
}
\eqn\efour{
A_{\theta}= {n \over r}v(r)
}
\noindent
where $r$\ and $\theta$ are polar coordinates on the 2-plane containing the
vortex; we will consider $n=2$. As usual we have  $A_0= A_3 = A_r = 0$.
We parametrize a perturbation about this solution as:
\eqn\esix{
\phi(r,\theta) = \Phi(r)exp(2i\theta) + \eta \overline {\phi}(r,\theta)
\ }
\eqn\esixa{
A_{\theta}= {2 \over r} (v(r) + \eta \overline {v}(r,\theta)).
\ }
\eqn\esixb{
A_{r}=\eta \overline {w}(r,\theta).
\ }
where $\eta <<1$ is a small parameter.

In particular, when $\alpha = 0$, perturbations of the form
\eqn\eseven{
\overline {\phi}(r,\theta) = \omega \Phi(r)
\ }
\eqn\esevena{
A_i = A_i^{(0)} + \eta \epsilon _{ij} \partial_j \omega.
\ }
where $i,j = 1,2$ and $A_i^{(0)}$ is the cylindrically symmetric solution of
the equation of motion,  produce a zero mode if $\omega$ satisfies
$\Delta \omega = 2 \Phi ^2 \omega$.
These zero modes of the energy may be resolved into eigenfunctions of the
azimuthal angular momentum operator, so we may label the solutions as
$\omega = \omega_m(r) cos(m \theta)$.
For $m=1$ the perturbation is nothing but a translation of the vortex in
its own plane, while modes with  $m>1$ correspond to distortions
of the string core. For finiteness of energy one must also require that
$m \le \nu_\phi$ where $\nu_\phi$\ is the winding number of $\phi$.

For $\alpha > 0$, one can generalize these perturbations so that they
reduce to the above form when $\alpha = 0$. The translational zero mode
exists in both cases; however, modes with $m>1$ now {\it decrease} the energy,
signalling the instability of the symmetric solution.

We now consider this analysis for the combined ($\phi, \chi$) system.
Our boundary conditions require the winding numbers of $\phi$
and $\chi$ to be $\nu_\phi=2$ and $\nu_\chi=1$ respectively.
The $(\phi , A)$ sector is
identical to the system in \Bog\ with $\nu_\phi= 2$. It is sufficient
to consider the $m = 2$  perturbation in the $(\phi , A)$ sector without
disturbing the configuration of the scalar field $\chi$, since this only
underestimates the extent of instability.  Let us first
expand the change in the vortex energy due to this perturbation in a
power series in $\alpha$ and $\gamma$ for  fixed $\lambda _2$,
keeping $\kappa=0$ for now.  We separate the change in
energy into $\gamma$-dependent and $\gamma$-independent parts:
\eqn\eten{
\delta {E} = \delta {E}_1 + \delta {E}_2
\ }
where the first term on the R.H.S is $\gamma$-independent. In fact,
$\delta {E}_1$ is equal to $\delta {E}$ for the system in \Bog.
$\delta {E}_1$ vanishes for $\alpha =0$, so we can expand it in
a power series in $\alpha$;
\eqn\elvn{
\delta {E}_1 = \alpha ^{c_1} (K_0 + K_1 \alpha + ....)
\ }
where the coefficients are all independent of the parameters of the
$\chi$ sector.
$\delta  {E}_2$ is the $\gamma$-dependent part; this part is regular
(at zero) in both $\gamma$ and $\alpha$ (regularity in $\gamma$
follows from the fact that $\delta {E}_2$ is zero for $\gamma =0$).
Therefore it can be expanded as
\eqn\etwel{
\delta {E}_2 = \gamma ^{c_2} (L_0(\alpha) + L_1(\alpha)\gamma +...)
\ }
All the $\alpha$-dependent coefficients of the expansion are regular in
$\alpha$.

We have already seen that $\delta {E}_1 < 0$ for $\alpha>0$
(and hence $K_0 < 0$), while earlier we argued that for
$\gamma > 0$, $\delta {E}_2 > 0$.  There is then
a curve in the $(\gamma,\ \alpha)$ plane, passing through
$\gamma = \alpha = 0$ and extending into the positive quadrant, along
which there is a zero mode, \ie, $\delta {E}_1 + \delta {E}_2 = 0$.
This curve separates the positive quadrant in the $\gamma$, $\alpha$
plane into two parts with opposite signs for $\delta {E}$, and
represents the line of marginal stability for the cylindrically symmetric
solution.  Since $\delta {E}$ is negative in the part
containing the axis $\gamma = 0$,  for every $\alpha>0$ there is a
range of values of $\gamma$ for which any cylindrically symmetric
solution to the equations of motion is unstable. However, it is clear
that because of the field $\chi$ the $\nu_\phi=2$ string can not split
completely into two well-separated strings, since isolated $\nu_\phi=1$
strings do not have finite energy per unit length.
The ground state in this region of parameter
space then must be a localized but asymmetric cosmic string.

This proves the existence of binary strings in this model when
$\kappa=0$.  The restriction to $\kappa =0$\ is not necessary, since
just as we found a range of binary-string solutions in the
$\alpha - \gamma$ plane, we can continue this region of solutions
into the third parameter-dimension, corresponding to
the parameter $\kappa$.  Since the contribution of this term is
nonsingular, for small values of
$\kappa$\ the cylindrical solution is still unstable.  There is then no
fine-tuning necessary to obtain this class of solutions, provided only
that the desired pattern of symmetry breaking takes place.  In
particular, if we assume that the second symmetry breaking takes
place at a lower energy scale, one does not expect the contributions
from the lighter scalar to affect the question of whether the $\nu_\phi =2$
strings are stable, and so we would expect the binary string to be
favored over a wide range of parameters, so long as $\alpha>0$.

(Note that we could have added to the Lagrangian a term
($\phi^*\chi^2 + c.c.$)\ which removes the
global symmetry allowing independent rotations of $\phi$\ and $\chi$; it
can be similarly dealt with, and will be discussed further at the end
of this section.)

Our second example of binary strings confined by a long-range force is
provided by a broken $U(1) \times \tilde {U}(1)$ theory. This time we
consider three charged scalar fields $\phi, \chi$ and $\tau$ with
charges (1,0), (0,1) and ($\demi,\ \demi$) respectively.

The lagrangian is

\eqn\etwon{
{ L}= {1\over 2}|D\phi|^2 - {1\over 4}F_{\mu\nu}F^{\mu\nu} +
{1\over 2}|D\chi|^2 -{1\over 4}G_{\mu\nu}G^{\mu\nu} +{1\over
2}|D\tau|^2- V.
\ }
\noindent
F and G are the kinetic terms for the two gauge fields $A_{\mu}$ and
$B_{\mu}$. We also have,
\eqn\ett{
V = \lambda_1 (|\phi|^2 - v_1^2)^2 +  \lambda_2 (|\chi|^2 -  v_2^2)^2
+ \lambda_3(|\tau|^2 - v_3^2)^2 + \alpha (|\phi|^2 - v_1^2)(|\chi|^2 -
 v_2^2) +
\ }
$$
\kappa_1 (|\phi|^2 -v_1^2)(|\tau|^2 -  v_3^2) +
\kappa _2 (|\chi|^2 -  v_2^2)(|\tau| ^2 -  v_3^2).
$$
We parametrize the ratio of the \vev's by
\eqn\etta{
\zeta^2= {v_2^2 \over v_1^2};\ \
\gamma^2 = {v_3^2 \over v_1^2}.
}

 Once again we first restrict the parameter
space by putting $\kappa _1 = \kappa _2 = 0$.  The \vev's of $\phi$,
$\chi$ and $\tau$ are respectively $v_1$, $\zeta v_1$\ and $\gamma
v_1$. We expect to find stable solutions only for boundary conditions
for which the phase of each scalar field has an integer winding
number at large distances; a minimal choice gives the winding numbers
 1, -1 and 0, respectively.  With this choice of couplings and boundary
conditions, $\tau$ will couple only weakly to the vortex, so we
expect that $|\tau| \sim \gamma v_1$ everywhere.

Let us momentarily consider the limit $\alpha = \gamma = 0$; the $\tau$
sector contributes nothing to the vacuum energy, and the $\phi$ and
$\chi$ sectors decouple. Two independent strings are formed, one coming
from the winding of $\phi$ and the other from the winding of $\chi$.
Although these two vortices will each be cylindrically symmetric and
stable, the two-vortex state has a trivial zero-mode describing the
relative displacement of the two vortices.  This zero mode becomes an
instability when one allows $\alpha$ to be greater than zero while
keeping $\gamma = 0$, since now the vortex cores repel one another (with
a short-range force).  On the other hand, allowing $\gamma$ to be non
zero while keeping $\alpha = 0$ will also eliminate the zero mode, as
the half-integer charge of $\tau$\ confines the singly-charged vortices
in pairs with $Q_A+Q_B = 0\ mod\ 2$.  For $\alpha = 0$ the ground state
will be cylindrically symmetric, but taking $\alpha$ finite and
increasing $\gamma $ from zero will clearly lead to an asymmetric ground
state, since in that case the confining force is long-range but small.
One can now argue (as in the first example of this section) that
allowing both $\alpha$ and $\gamma $ to be greater than zero one will
obtain a curve of marginal stability in the positive quadrant of the
$\alpha, \gamma$ plane; the part of this quadrant between the axis
$\gamma = 0$ and the curve of marginal stability is the region of
parameter space for which doubly-charged, cylindrically-symmetric
vortices are unstable to formation of confined pairs of singly-charged
vortices, \ie\ binary strings.

Our initial restriction on the other scalar self-couplings is once again
inessential.  One can now add two more dimensions (corresponding to
$\kappa _1$ and $\kappa _2$) to our parameter space, and the region of
parameter space for which the binary string will be the ground state
will be determined by some inequalities among the different
contributions to the kinetic and potential energies.
Since even a modest hierarchy among the \vev's $v_i$\ is sufficient to
ensure that these inequalities pose little restraint on the parameter
space, we see that for these models the binary string is technically natural
and even generic, in the same sense that type I and type II vortices both
occur in the phenomenology of superconductors.

\subsec{Goldstone modes and the binding force}

An interesting feature of the above models is the presence of a
Goldstone boson in the broken phase. In both models we have omitted a
scalar self-coupling which links the phase rotations of the different
fields, and as a result there is an extra U(1) symmetry which is not
gauged and is spontaneously broken along with the gauged U(1)'s.  This
extra global symmetry will be preserved by renormalization if originally
present (unless, \eg, there are fermions and the global symmetry is
anomalous), and could arise as a result of discrete symmetries in a more
fundamental theory, or through the absence of suitable renormalizable
couplings in a unified model.  In realistic models there are no massless
scalars, and so either the symmetry-breaking coupling is present,
or else perhaps the extra U(1) can be gauged. When the symmetry-breaking
coupling is small enough the binary strings will exist by our earlier
continuity argument.

Let us define what we mean by the binding force.
Consider quasi-statically separating the components
of the string transversely along their length.
Although these strings cannot be divided into their constituent strings,
there will be deformations that stretch the binary string
into two string-like boundaries of a domain-wall-like structure.
Thus, consider a one-parameter family of deformations of the
binary vortex in the first model, where the field
$\phi$ is constrained to vanish at two different points
in the transverse plane at a distance $R$.
Finding the minimum energy solution to the equations of motion with this
constraint, we define the binding force as (-) the derivative of
the energy with respect to the parameter $R$.

The fields forming the domain wall are massive in the true vacuum and the
domain wall has a thickness $h$ of order ${1\over m}$. The
energy of the domain wall is then a linear function of its area when
the area is large compared to the thickness. For binary
strings this translates to a binding force that is constant when $R$ is
large compared to the thickness of the domain wall.
When the `domain wall' is relatively thick ($h>>R$)
the energy grows less than linearly with the separation $R$,
perhaps logarithmically. In fact, for small $R$\ the
fields in the region between the strings generally do not resemble
a domain-wall.  This small-separation case is quite
typical of grand-unified models of binary strings, since the binding
force and the short-distance repulsion may come from vastly different
energy scales.  Moreover, the nature of the binding force  affects
the action describing the motion of the binary string\strmech.

If the Goldstone mode is present, again the restriction
to values of $R$ much larger than $h$ is insufficient.
Nevertheless, by its nature the Goldstone mode is derivatively coupled
and for large enough $R$ the contribution from the massive fields
to the domain wall energy will dominate the effects of the Goldstone boson
due to their linear increase with $R$.
Because of the strong changes in the fields near the string cores,
the Goldstone boson may couple strongly to the motion and
internal excitation of the strings.

\newsec{Molecular binaries}

Finally we consider models in which the attractive force between the two
vortices or strings is short-ranged, and the existence of binary strings
is due to a minimum in the vortex-vortex potential.  The energy required
to separate two vortices is then finite, but for strings of cosmological
length, separation is energetically impossible.
Consider a $U(1)\times{\tilde U(1)}$ gauge theory with complex scalar
fields $\phi_1 , \phi_2$ and a neutral scalar $\sigma$, with charges
(1,0), (0,1) and (0,0) respectively. The Lagrangian is,
\eqn\Lmob{
{\it L}= {1\over 2}|D\phi_1|^2 - {1\over 4}F_{\mu\nu}F^{\mu\nu} +
{1\over 2}|D\phi_2|^2 - {1\over 4}G_{\mu\nu}G^{\mu\nu} +
{1\over 2}|\partial\sigma|^2 - V(\phi_1,\phi_2,\sigma),
}
\noindent
where $F$\ and $G$\ are the field strengths of $A$\ and $B$\
respectively, $D_\mu\phi_1=(\partial_\mu+ieA_\mu)\phi_1$, and
$D_\mu\phi_2=(\partial_\mu+ieB_\mu)\phi_2$. The potential is
\eqn\mbpotl{
V(\phi_1,\phi_2,\sigma)= \lambda_1(|\phi_1|^2-\mu_1^2)^2 +
\lambda_2\sigma^4+ m^2\sigma^2- \lambda_3(\mu_1^2-|\phi_1|^2)\sigma^2 +
}
$$\lambda_4(|\phi_2|^2-\mu_2^2)^2 -\lambda_5(\mu_2^2-|\phi_2|^2)
(\mu_1^2-|\phi_1|^2)+
\lambda_6(\mu_2^2-|\phi_2|^2)\sigma^2.
$$
\noindent
Note that the off-diagonal quartic couplings $\lambda_5$ and $\lambda_6$
have the `wrong sign'; for a stable vacuum to exist this potential must
be bounded from below. If the masses and coupling constants are all
positive, a sufficiency condition for the positivity of the energy is
easy to write: defining
 $V=V^\prime + m^2\sigma^2$ we see that the positivity of $V^\prime$ is
sufficient to guarantee the positivity of $V$.
\smallskip
Now define $x_1=\mu_1^2-\phi_1^2$, $x_2= \sigma^2$,
$x_3=\mu_2^2-\phi_2^2$, and in a matrix notation we can write
$V\prime = X^T \Lambda X$ ; where $X = (x_1 x_2 x_3)$ and
\eqn\matpotl{\Lambda =
\left( \matrix {\lambda_1 & -\lambda_3/2 & -\lambda_5/2 \cr -\lambda_3/2
& \lambda_2 & \lambda_6/2 \cr -\lambda_5/2 & \lambda_6/2 & \lambda_4
\cr} \right).}
\smallskip

$V\prime$ is positive definite if $\Lambda$ has no negative eigenvalues,
which implies that the coefficients of the characteristic equation of
$\Lambda$\ must all be non-negative.  This gives rise to some
inequalities involving the couplings:
\eqn\inone{Tr\Lambda = \lambda_1 + \lambda_2 + \lambda_4 >= 0,}
\eqn\intwo{  det \Lambda = 4\lambda_1\lambda_2\lambda_4 +
\lambda_3\lambda_5\lambda_6
-\lambda_1\lambda_6^2 - \lambda_2\lambda_5^2 -\lambda_3^2\lambda_4 >=
0,}
\eqn\inthr{4[\lambda_1\lambda_2 + \lambda_2\lambda_4 + \lambda_4\lambda_1] -
\lambda_3^2 - \lambda_5^2 - \lambda_6^2 >= 0.}
\smallskip
When these conditions are satisfied the potential energy can be written
as a sum of squares by diagonalizing $\Lambda$ so it is explicitly
non-negative, furthermore, we have arranged that $V=0$ in the desired vacuum.
We will assume that all the $\lambda$'s are positive thus satisfying
\inone at once. We shall come back to the other conditions on the
$\lambda$'s later, when the other requirements are known.

First let us note that with this Higgs potential, the
$U(1)\times{\overline U(1)}$ symmetry is spontaneously broken; the
vacuum expectation value of the symmetry breaking Higgs fields are
$|\phi_{1,vev}|=\mu_1 $\ and $|\phi_{2,vev}|=\mu_2$; and the \vev\ of the
singlet field $\sigma$ is zero in the true vacuum.  The  manifold of
vacua has nontrivial topology;
$\pi_1(U(1)\times{\tilde U(1)})=Z\times {\tilde Z}$ implies the
existence of two kinds of strings: those with a flux of A and
those with a flux of B.  We shall call them type A and type B
vortices, respectively. These can occur {\it independently}, so the
binding which occurs is through short-range forces only.
As we shall see, the vortex-vortex binding can
occur between vortices of different types.

Vortices of circulation one in either A or B alone are always stable.
We are interested in the case of a pair of parallel A and B type strings
with a \vev\ of $\sigma$ in the core of the A type string.
Let us review the conditions for which $\sigma$ would get a \vev\ in
the core of the A type string.
The terms in the potential involving $\sigma$ are,
\eqn\vsig{
V(\sigma)=\lambda_2\sigma^4+m^2\sigma^2-\lambda_3(\mu_1^2-\phi_1^2)\sigma^2.}
\noindent
At the core of A type string we can write a local potential for $\sigma$
with other fields replaced by their values at $r=0$,
$\phi_{1}=0,\ \phi_{2}=\mu_2$:
\eqn\vcsig{ V(\sigma)=
\lambda_2\sigma^4+m^2\sigma^2-\lambda_3(\mu_1^2)\sigma^2.}
\noindent
There is a non trivial minimum of this potential at
$\sigma^2_{vev}={\lambda_3\mu_1^2-m^2\over 2\lambda_2}$, when
$\lambda_3 \mu_1^2 > m^2$.

We can write a similar potential at the core of a string of type B;
however, the sign of the term with $\lambda_6$ in the potential has been
chosen so that the minimum of the `core' potential is now at
$\sigma_{vev}=0$.  The true configuration of the $\sigma$ field inside
the A vortex is of course dictated by a balance between the
gradient and potential contributions to the energy. However,
when the condition
\eqn\infor{\lambda_3 \mu_1^2>m^2}
\noindent
is fulfilled, a nonzero value of $\sigma_{vev}$ is preferred at the core
of the A type strings because of the gain in potential energy.

The conditions for obtaining a condensate of a `spectator' scalar field
were examined in \SCS; let us briefly recapitulate.
We begin with a static, cylindrically-symmetric vortex solution
(Nielsen-Olesen vortex) for $\phi_1$ and the gauge field $A_{\mu}$,
neglecting $\phi_2$\ for now.  We wish to see
if $\sigma =0$ is indeed the vacuum state;
we examine small oscillations of $\sigma$ around the background $\sigma =0$,
$ \sigma = \sigma_0(r) e^{i \omega t}$.
The linearized equation of motion of $\sigma$ is,
\eqn\linsig
{
- {\partial ^2 \sigma_0(r) \over \partial r^2}
- {1 \over r} {\partial \sigma_0(r) \over \partial r}
- P\sigma_0(r) = \omega ^2 \sigma_0(r)
}
When there are solutions of \linsig with $\omega^2 < 0$,
$\sigma=0$ is an unstable solution of the classical equations \SCS.
When $m^2=0$, the potential energy of this 2d Schroedinger equation,
$P\equiv\lambda_3(\mu_1^2 - \phi_1^2)$ is everywhere negative when the
background is the Nielsen-Olesen vortex solution $\phi_1(N.O)$.
The argument in \SCS then proceeds that, since a negative definite
potential in two dimensions always has a bound state,
$\omega^2$ has a negative value for $m^2=0$;
by continuity, then, in the neighborhood of $m^2=0$,
$\sigma =0$ is an unstable solution
which relaxes to a non zero \vev\ for $\sigma$.
Indeed, $\sigma$ develops a nonzero \vev\ whenever $m^2 < \mu^2$,
where $-\mu^2$ is the most negative eigenvalue for $\omega^2$
in equation \inthr\ with $m=0$.

Although we cannot express $\mu$\ directly in terms of the
parameters of the theory, it can be easily approximated by
doing a simple variational calculation, or even more simply, estimated
by making an ansatz for the zero mode, parametrizing the shape
of the vortex by elementary functions.
Choosing the latter method (taking the ansatz for $\sigma_0$ to
be given by the difference of two exponentials) we get
$$
\omega^2 \sim 2.4k^2 - 0.6 \lambda_3 \mu_1^2.
$$
where $k\equiv \sqrt{2\lambda_1}\mu_1$.
The first term on the {\it RHS} is from the kinetic energy and the
second is from the potential energy of the $\sigma$ condensate.
There are two things to note:
(a) This value of $\omega^2$ underestimates the actual value of $\mu^2$.
However, we can also obtain an upper bound;
for a vortex with $|\phi_1(r)|$ that monotonically increases
from $\phi_1=0$ at $r=0$ to $\phi_1=\mu_1$ at $r=\infty$, we must have
$\mu^2 \le \lambda_3 \mu_1^2$.
(b) The value of $\mu^2$ is an increasing function of the
range of the potential, when the strength of the potential is held
constant. This is because the potential energy term $P(r)$
increases in magnitude point-wise as the range of the
potential is increased.  In our case the range of the
potential is $1/k$, so $\mu^2$ is an increasing function of k.

We want to show that there is a range of parameters for which
(i) $\sigma$ develops a \vev\ in the core of the A type strings;
and (ii) the asymptotic forms of the $\phi_1$ and $\sigma$ fields is
$|\phi_1 |(r) \sim \mu_1 -\phi_1e^{-kr}$
and $|\sigma|(r) \sim \sigma_0 e^{- m r}$.
(i) will be true if the condition $m^2 \le \mu^2 $ is satisfied.
Since $\mu^2 \le \lambda_3 \mu_1^2 $ we can define a vortex
parameter $f_{\phi_1}$ (depending on the vortex field background
$\phi_1$) by
\eqn\vortpar{
|\mu^2|_{\phi_1}= f_{\phi_1} \mu_1^2 \lambda_3,
}
with $0 < f_{\phi_1} <=1.$

Condition (ii) for $|\sigma|(r) $ simply follows from the
linearized classical equations in the asymptotic region.
Similarly we obtain the correct asymptotic behavior for $\phi_1$ if
$m^2>8 \lambda_1 \mu_1^2$ and $e^2>2\lambda_1^2$.
For consistency with (i) we then must have,
\eqn\8 {
f_{\phi_1(\lambda_1)} \lambda_3>8\lambda_1.
}
For given values of $\mu_1^2$ and $\lambda_3$ there exist a range of
values for $\lambda_1$ for which this consistency condition is
satisfied. This is quite natural since the width of the Nielsen-Olesen
vortices increases as $\lambda_1 \rightarrow 0$; in that limit,
$\mu^2$ tends to its maximum value:  $\mu^2 \rightarrow \lambda_3 \mu_1^2$.
Thus, given any $\lambda_3 > 0$, there is a range of values of
$\lambda_1$ extending to zero satisfying
$f_{\phi(N.O)} \lambda_3 > 8 \lambda_1$.

Having established the asymptotic behavior of the $\phi_1$ and
$\sigma$ fields when $f_{\phi_1} \lambda_3 > \lambda_1$ we now examine
the behavior of a B-type vortex in the background of an A-type vortex,
which as we see has a $\sigma$ condensate in the core. The scale that
determines the width of the B type vortex
is the mass of the $\phi_2$ field in the true vacuum,
width(B) $ \sim { 1 \over {\sqrt {\lambda_2 \mu_2^2}}}$.
We assume that the $(\phi_1 , \sigma)$ and the $\phi_2$ sectors have a
large difference in mass scales, $\mu_2^2 >> \mu_1^2 \sim m^2$.
This would be natural for $\lambda_5=\lambda_6=0$, when the two sectors
are decoupled. In a GUT version this hierarchy would result from
symmetry breaking at different scales; in the present case, it does not
need to be very large. In this simplified model
we assume $\lambda_5$, $\lambda_6$
to be relatively small:
$\lambda_2, \lambda_3, \lambda_4 >> \lambda_5, \lambda_6$.

The large mass scale of the B type vortex ensures that the B type
strings are very thin compared to the A type ones. We will then
approximate the `shape' of the B vortex effectively
by a delta function, with the field $\phi_2$ dropping abruptly to zero
from its true \vev\ $\mu_2$ at the core of the vortex, for the purpose
of considering the interaction of the vortices.
If we bring such a (thin) B-type string near an A-type string, while
keeping them parallel, the structure of the vortices will not be
changed appreciably by their mutual interaction. Indeed the interaction
is purely through quartic Higgs couplings of
$\lambda_5$ and $\lambda_6$, which can be treated as a perturbation.

We can now calculate the interaction energy  per unit length
of a pair of parallel A and B type strings,
$V_{int}={\mu^\prime}^2
[\lambda_6\sigma^2(r^\prime)-\lambda_5(\mu_1^2-\phi_1^2(r^\prime))]$,
where ${\mu^\prime}\sim \mu_2^2$ is determined by integrating the
interaction energy over the cross section of the thin B-type vortex,
and $r^\prime$ is the position of the B-string
relative to the core of the A-string.
\smallskip
When the B string is far away from the core of the A string we can use
the asymptotic forms of the $\phi_1$ and $\sigma$ fields:
$$
|\phi_1(r)| \simeq \mu_1 -\phi_1 e^{{\sqrt {-8\lambda_1}} \mu_1(r)};\
$$
$$
| \sigma(r)| \simeq \sigma e^{-m r};
$$
$$
V_{int}=\mu_2[\lambda_6 \sigma^2e^{-2m r}-2
\lambda_5\mu_1 \phi e^{-{\sqrt8}\lambda_1\mu_1(r)}] + h.o.
$$
At $r\rightarrow\infty$, ${\partial V_{int} \over \partial r}$
is positive ($m^2 > 2\lambda_1 \mu_1^2$), giving
an attractive force between the vortices. This force decreases to zero
and changes sign at the equilibrium point $r=r_{eq}$ which is given by
$2\lambda_6\sigma ^2 m e^{-2 \omega r_{eq}}-2\lambda_5\mu_1
\phi_1 e^{-{\sqrt{\lambda_1\mu_1(r_{eq})}}}
+ h.o. = 0$.

The vortex-vortex interaction leads to a equilibrium
separation of the strings, $r_{eq}$. By choosing an appropriate ratio of
the constants $\lambda_5$ and $\lambda_6$, $r_{eq}$ can be made to
fall within the range of validity of our approximations.
In the present case, $\lambda_6$ must be much larger than $\lambda_5$.
(That is, we have balanced a somewhat-longer range attraction with
a stronger, but shorter ranged repulsion.)
This will obviously lead to a binding energy per unit length
between the two strings that is several orders of magnitude smaller
than the mass per unit length of the binary string, since
the attractive force is proportional to $\lambda_5$.
To get a larger binding energy one could increase $\lambda_5$ (and then
$\lambda_6$ as well), in which case one must replace the asymptotic forms
of the $\phi_1$ and $\sigma$ fields by the exact classical solution,
since $r_{eq}$ moves into the core of the A-string. There is
{\it a priori} no reason that the qualitative
features would change drastically for more strongly bound strings;
there will be some range of couplings for which the binary-string
ground state is obtained, with the limit on the strong-binding side
being set by the fact that as the couplings get stronger and the
strings are more closely bound, it becomes energetically favorable
to deform the A-string and obtain a cylindrically-symmetric configuration.
In any case we have an existence proof of vortex-vortex binding at
finite separation in this model, and presumably others of the
same class can be found.

(Returning to the constraints on our scalar potential, the nontrivial
conditions \intwo, \inthr\ and \8 in the above model are met for a wide
range of  (positive) parameters $\lambda$. For instance when
$\lambda_5$ and $\lambda_6$ are sufficiently small the conditions
essentially reduce to
$$
4\lambda_1 \lambda_2 \lambda_4 -\lambda_3^2 \lambda_4 > \delta,
$$
$$
4[\lambda_1 \lambda_2 + \lambda_2\lambda_4 + \lambda_4\lambda_1] -
\lambda_3^2> \delta,
$$
$$
f_\phi\lambda_3 >8 \lambda_1,
$$
where $\delta$ is a small positive parameter.
If $\lambda_1 = \epsilon f_{\phi} \lambda_3$, with $\epsilon <1$, then
conditions
\intwo, \inthr and \8 are easily satisfied by any partial ordering
$\lambda_2, \lambda_4 >\lambda_3$ satisfying
$4\lambda_2>\lambda_3/{\epsilon f_{\phi}}$.)

To compare this with the previously-constructed models, the most
significant difference is the finite binding-energy per unit length,
which is much smaller than the string tension.
The individual strings in this binary are independently stable; if they
were initially separate, they would be very unlikely to combine
because the interaction energy between two intersecting segments
of string could not compete with the kinetic energy of the rest of
the string under normal circumstances. On the other hand, if such
a binary string were formed it would be very difficult
to dissociate, since a macroscopic string would have
a very large binding energy, and it would be necessary
to transfer sufficient energy to the whole string, rather
than an individual segment.

\newsec{\bf GUT considerations}

Binary strings of the `confining' type will occur in grand-unified
theories provided two conditions are satisfied: the sequence of
symmetry-breaking phase transitions gives rise to cosmic strings and
broken discrete symmetries in the necessary order, and the coupling
constants are such that multiple-winding strings are unstable.  The
first of these conditions is the more stringent one.  The question of
whether the pattern of symmetry breaking is appropriate depends on the
representations of the symmetry-breaking order parameters (\eg Higgs
fields) and the scales at which the symmetries are broken.  (Note that
although we will be describing the symmetry breaking in terms of
elementary scalar fields, there is no reason why other symmetry-breaking
mechanisms, such as fermion condensates or Wilson lines, could not play
a role).

The first type of confining binary, in which two similar strings are
bound together by a (vestigial) domain wall, is the most generic, since
it could occur in any grand-unified model in which strings are first
formed and at a later phase transition bound domain walls.  It may be
surprising that a stable remnant persists in this case; however, we
observe that the orientation of the domain wall is not gauge-invariant,
but changes direction under the $Z_2$\ discrete gauge symmetry
spontaneously broken by the lighter scalar field, so the string can
assume either orientation as the boundary of the domain wall.  (In terms
of the holonomy of the gauge field, the same group element, (-1) in the
representation of the charge-1 field, is obtained in passing around the
string in either direction, and likewise passing in either direction
through the domain wall.)  In the next section, we remark on how this
circumstance allows (n=2) strings, whether binary or not, to survive
the phase transition in which the domain walls form.

This type of confining behavior can occur in a well-known model of
SO(10) grand-unification described by \KLS\ in their analysis
of string-bounded domain walls.  In that symmetry breaking sequence,
\eqn\KLSgut{
Spin(10) \overarrow{\bf 54}
Spin(6) \otimes Spin(4) \overarrow{\bf 126}
SU(3) \otimes SU(2) \otimes U(1) \overarrow{\bf 10}
SU(3) \otimes  U(1)
}
cosmic strings are formed (of the `Alice string' variety
\ref\Alicea{A.S. Schwarz, {\it Nucl.Phys.} {\bf B208} (1982) 141;
A.S. Schwarz and Yu.S. Tyupkin, {\it Nucl.Phys.} {\bf B209} (1982) 427.}
\ref\Aliceb{M.G. Alford, K. Benson, S. Coleman, J. March-Russell and
F. Wilczek,{\it Phys.Rev.Lett.} {\bf 64} (1990) 1632, (E){\bf 65}
(1990) 668;
J. Preskill and L. Krauss, {\it Nucl.Phys.} {\bf B341} (1990) 50.})
at the first phase transition; the $ {\bf 126}$ \vev\ is not
$\cal C$-invariant, and so at the second phase transition the strings
bound domain walls.  The surviving strings will be of the binary or
the $n=2$ type, depending on the relative sizes of the scalar and
gauge couplings, as previously discussed.  Other examples may be
similarly constructed.

Incidently, this same model (in a different region of parameter space)
might exhibit the `molecular' variety of binary string.  At the phase
transition in which the domain walls are formed, there are
independently formed \KLS\ stable cosmic strings (corresponding to
the discrete gauge symmetry $(-1)^{F_{10}}\in Spin(10)$).  They
might be expected to be comparable in abundance to the remnant $n=2$
strings remaining after the domain walls have disappeared.  One
could then extend the scalar sector of this model so that the situation
of section 4 is realized, that is, a weak attraction at long distances
between the strings, and a short-distance repulsion, due to a scalar
condensate in the core.  In view of the somewhat dubious cosmology
of this type of defect, we will not develop such a model in detail.

If we consider grand-unification based on the $E_6$ gauge group, there
are more possibilities for the unbroken discrete gauge symmetries
and hence a variety of string solutions.  So, for instance, if at
the highest scale we break $E_6$ to $SU(5)\otimes U(1)\otimes U(1)$
via one or more adjoints (${\bf 78}$'s), we can obtain independent
cosmic strings at subsequent stages of symmetry breaking, either
in combination with or independent of $SU(5)$ breaking.  These could
furnish examples of any of the varieties of binary string we have
described.

We will present one concrete example along a somewhat different line.
Consider an $E_6$ GUT model, with a symmetry-breaking sequence
\eqn\Esix{
E_6 \overarrow{\bf 27} SO(10)\otimes Z_4
\overarrow{\bf 351^\prime} SU(5)\times Z_2\otimes Z_2
\overarrow{\bf 27} SU(5)\otimes Z_2
\rightarrow \dots
}
(we leave the subsequent symmetry breaking to the reader's imagination.)
The representations of the scalar \vev's are, respectively, a {\bf 27}
$\phi$, a ${\bf 351^\prime}\ \chi$, and another {\bf 27} $\tau$.  (Note
that the
${\bf 351'}$ is in the symmetric product of two $\bar {\bf 27}$'s, and
the direction of the \vev\  is chosen to lie in the {\bf 126} direction
in its $SO(10)$ decomposition, so the $Z_2$ string is similar to those
in $SO(10)$ GUT models).  Our nicknames for the scalars is meant to
suggest the similarity to the second model of section 2; indeed,
the only difference is that in this case the discrete symmetry group
has a different breaking pattern, $Z_4 \otimes U(1)\rightarrow
Z_2\otimes Z_2 \rightarrow Z_2$.  Because of the $Z_4$\ breaking the
composite strings may form at the second phase transition, as well
as at the third.

Finally, we once again emphasize that the point of this exercise is
not to find compelling grand-unified models with binary strings,
although the first example would do for that, but rather to
show  that in non-minimal unification the conditions for the
existence of composite strings are easily satisfied.

\newsec{\bf Formation and Evolution of Binary Cosmic Strings}

The formation of binary cosmic strings begins by the Kibble
mechanism \ref\KibMech{T.W.B. Kibble, {\it J.Phys.} {\bf A9} (1976) 1387}:
the Higgs scalar develops an expectation value below some critical
temperature $T_c$, which is disordered on scales larger than some
correlation length $\xi_\phi$\ determined by the kinetics of the phase
transition.  As the scalar field relaxes, defects are formed at
the boundaries of domains with different orientations of the condensate.
At lower temperatures, the scalar field fluctuations are
suppressed and the network of defects begins to evolve.

The details of binary string formation depend on the mechanism
responsible for the internal structure and the evolution subsequent to
the initial formation of the strings.  In the case of confined binary
strings domain walls are formed at a later phase transitions when one of
the discrete gauge symmetries is broken.  Since the strings arising from
the earlier symmetry breaking are charged under the discrete symmetry,
they will become boundaries of domain walls.  As the network of domain
walls forms in the presence of the strings the relative orientations of
different components of the boundary of a given wall is determined by
chance. Then, as the network evolves, some components of the boundary
will meet as the domain wall shrinks; those that have opposite
orientations will annihilate, while those segments of string that have
the same orientation will form linear defects with winding number $n=2$.
These will be binary strings, when the couplings are such that the cores
are repulsive; otherwise they will form metastable,
cylindrically-symmetric $n=2$\ strings.  (In both cases the strings are
unstable to decay by $Z_2$ monopole-pair production
\ref\strdk{A. Vilenkin, {\it Nucl.Phys.}{\bf B196} (1982) 240},
through a greatly suppressed tunneling process; because
of the separation of the cores the rate will be even more heavily
suppressed in the case of binary strings.)

The situation is similar with the other confining model, in which two
different types of string are bound together at the
discrete-symmetry-breaking phase transition.  Here, however, the
domain-wall can end with either orientation, on either of the
types of string.  Generally we may assume one of these
strings is much lighter and  more abundant than the heavier ones
initially formed.  It is then reasonable to expect that while many of
the lighter strings may annihilate, nearly all of the heavier strings
should be paired up with the more abundant lighter strings.  (Note that
in this case if the couplings do not make the cores mutually repulsive,
the final state will have strings with a gauge-field flux being a
composite of that comprising the initial strings.)

The circumstances under which the `van der Waals' or `molecular' strings
may be formed cosmologically are quite unclear.  There may be
circumstances under which preexisting strings from an earlier phase
transition could serve as nucleation sites for the subsequent transition
or otherwise influence the ordering of the scalar field condensate, in
such a fashion that the binary strings are formed together.  Otherwise,
since the separate strings are stable and the interaction in this case
is short-ranged, there is vanishing probability of forming binary
strings when the component strings are formed separately.

After formation the evolution of a network of binary cosmic strings is
essentially the same as for ordinary cosmic strings, since at large
distances the string mechanics is dominated by the string tension (which
is nearly the sum of the tension for each component).  One difference is
that the initial network of strings formed by the collapse of the domain
walls may be somewhat different from the usual string network, but
differences of this type will presumably be washed out by the subsequent
evolution of the string network. Another effect (albeit a small one)
which should be noted is that the {\it effective} string tension and
mass on cosmological
scales \ref\effTm{A. Vilenkin, {\it Phys.Rev.} {\bf D41} (1990) 3038;
B. Carter, {\it Phys.Rev.} {\bf D41} (1990) 3869}
will be affected by the internal
degrees of freedom (twisting modes).

It is more interesting consider the coupling of the internal modes of
the binary cosmic string to the various epiphenomena of cosmic string
solutions: superconductivity, fermion zero modes, the kinks and cusps of
classical string evolution.  Some of these questions will be addressed
in the companion paper \strmech.

\medskip
\centerline{\bf Acknowledgments}

This work was supported by DOE contract \#DE-FG02-91ER40767.

\vfill\eject
\listrefs
\vfill\eject\end